\documentclass[a4paper,11pt]{article}
\pdfoutput=1
\usepackage{jheppub}

\usepackage{amssymb,amsmath,amsfonts}
\usepackage[normalem]{ulem}
\usepackage[utf8x]{inputenc}
\usepackage{slashed}
\usepackage{graphicx}
\usepackage{tabularx}
\usepackage{here}
\usepackage{color}
\usepackage{csquotes} 
\usepackage{comment}
\usepackage{mathrsfs}
\usepackage{float}
\usepackage{ascmac}
\usepackage{multirow}
\usepackage{longtable}
\usepackage{bm}
\usepackage{ulem}
\usepackage{caption}
\usepackage{nicematrix}
\usepackage{arydshln}

\usepackage[italicdiff]{physics}
\usepackage{tikz}
\usepackage[compat=1.1.0]{tikz-feynhand}
\def\slash#1{\not\!\!#1}

\makeatletter
\newcommand*\rel@kern[1]{\kern#1\dimexpr\macc@kerna}
\newcommand*\widebar[1]{%
  \begingroup
  \def\mathaccent##1##2{%
    \rel@kern{0.8}%
    \overline{\rel@kern{-0.8}\macc@nucleus\rel@kern{0.2}}%
    \rel@kern{-0.2}%
  }%
  \macc@depth\@ne
  \let\math@bgroup\@empty \let\math@egroup\macc@set@skewchar
  \mathsurround\z@ \frozen@everymath{\mathgroup\macc@group\relax}%
  \macc@set@skewchar\relax
  \let\mathaccentV\macc@nested@a
  \macc@nested@a\relax111{#1}%
  \endgroup
}
\makeatother

\numberwithin{equation}{section}

\preprint{
\begin{minipage}{5cm}
\small
\flushright
EPHOU-26-06\\ 
\end{minipage}}

\title{
More about modular symmetries and non-invertible properties in magnetized compactifications
}

\author{Tatsuo Kobayashi,} 
\author{Shuhei Miyamoto,} 
\author{Riku Nakano,} 
\author{Ryusei Nishida, \\ and } 
\author{Haruki Uchida}
\affiliation{
Department of Physics, Hokkaido University, Sapporo 060-0810, Japan\\
}

\emailAdd{kobayashi@particle.sci.hokudai.ac.jp}
\emailAdd{s-miyamoto@particle.sci.hokudai.ac.jp}
\emailAdd{r-nakano@particle.sci.hokudai.ac.jp}
\emailAdd{r-nishida@particle.sci.hokudai.ac.jp}
\emailAdd{haru-uchida@particle.sci.hokudai.ac.jp}

\abstract{ 
We study the modular symmetry in magnetized compactifications.
The zero-modes with different Scherk-Schwarz phases transform each other.
A generic model does not include modes with all the Scherk-Schwarz phases.
Incomplete multiplet representations appear.
Thus, the modular symmetry is violated as group-like symmetry.
However, the modular symmetry still controls coupling terms in those models.
Modular forms of the full symmetry appear as coupling constants.}

\makeatletter
\gdef\@fpheader{}
\makeatother

\begin{document}

\maketitle

\section{Introduction}

Symmetries are important in particle physics as well as other fields of physics.
Conventionally, group-like symmetries have been imposed in particle physics.
They control interactions of particles.
Abelian groups lead to coupling selection rules such that Abelian charged of particles must be conserved in allowed couplings.
That is the charge conservation law.
Non-Abelian group-like symmetries except Abelian parts require certain relations among couplings.

String theory on compact spaces and its effective field theory with extra dimensions lead to specific coupling selection rules, and they can be origins of symmetries, which are imposed in four-dimensional field theory.
Geometrical symmetries of compact spaces in string theory as well as higher dimensional field theory provide us with coupling selection rules in four-dimensional field theory.
Modular symmetry is one of important geometrical symmetries.
Different string states in heterotic orbifold models are transformed each other by the modular symmetry \cite{Ferrara:1989qb,Lerche:1989cs,Lauer:1989ax,Lauer:1990tm}.
Thus, the geometrical modular symmetry becomes a flavor symmetry in heterotic orbifold models and its low-energy effective field theory.
Calabi-Yau compatifications in heterotic string theory with the standard embedding also have modular flavor symmetries \cite{Strominger:1990pd,Candelas:1990pi,Ishiguro:2020nuf,Ishiguro:2021ccl}.
Also, modular flavor symmetries can be realized in magnetized D-brane models \cite{Kobayashi:2018rad,Kobayashi:2018bff,Ohki:2020bpo,Kikuchi:2020frp,Kikuchi:2020nxn,
Kikuchi:2021ogn,Almumin:2021fbk}. 
In these models, Yukawa couplings as well as other couplings are functions of moduli and transform under the modular symmetry.

Inspired by these aspects, modular flavor symmetries have been studied extensively as a bottom-up approach of model building and interesting results have been obtained.
(See Refs.~\cite{Feruglio:2017spp,Kobayashi:2018vbk,Penedo:2018nmg,Criado:2018thu,Kobayashi:2018scp,Novichkov:2018ovf,Novichkov:2018nkm,deAnda:2018ecu,Okada:2018yrn,Kobayashi:2018wkl,Novichkov:2018yse} for earlier works and Refs.~\cite{Kobayashi:2023zzc,Ding:2023htn} for reviews.)
In these modular flavor models, Yukawa coupling constants as well as other coupling constant are modular forms, which transform under the modular symmetry.
That is a novel point.
When we fix values of moduli, the modular flavor symmetries are broken.

Stringy theory on compact spaces lead not only to group-like  symmetries, but also to symmetries without group actions.
For example, strings in heterotic orbifold models are specified by their boundary conditions, $X(\sigma=\pi)=gX(\sigma = 0)$, where $g$ denotes a space group element.
However, $hX$ can satisfy the same boundary condition, 
$hX(\sigma=\pi)=ghX(\sigma = 0)$
where $h$ is another space group element, and that becomes 
$X(\sigma=\pi)=h^{-1}ghX(\sigma = 0)$.
Thus, each string state corresponds not to a group element, but to a conjugacy class, $h^{-1}gh$ \cite{Dixon:1986jc,Hamidi:1986vh,Dixon:1986qv}.
Multiplication rules of conjugacy classes are different from those of a group.\footnote{See Refs.~\cite{Kobayashi:1990mc,Kobayashi:1991rp,Kobayashi:2025ocp} for concrete examples of coupling selection rules in heterotic orbifold models and Ref.~\cite{Dong:2025jra} for various examples of multiplication rules of conjugacy classes. }
We write their elements, $x,y,z$, and their multiplications rules can be written by the fusion algebra, $xy=\sum_N N^z_{xy}z$.
The element $z$ in the right hand side is unique in group theory, but not always unique in multiplication rules of conjugacy classes, where one cannot define inverse elements.
That leads to non-invertible selection rules.

A similar behavior can be realized in magnetized compactifications.
There appear $M$ zero modes in magnetized D-brane models on $T^2$ with the magnetic flux $M$ \cite{Cremades:2004wa}, which are denoted by $\psi^k$ with $k=0,1,\cdots, M-1$.
Low-energy effective field theory has a $\mathbb{Z}_M$ symmetry, which transform the zero-modes as $\psi^k \to e^{2\pi ik/M}\psi^k$ \cite{Abe:2009vi,Berasaluce-Gonzalez:2012abm,Marchesano:2013ega}.
We consider the $T^2/\mathbb{Z}_2$ orbifold, where zero-modes are written by 
$\psi^k_{T^2/\mathbb{Z}_2}=\psi^k + \psi^{M-k}$ \cite{Abe:2008fi}.
The $\mathbb{Z}_M$ symmetry is broken on $T^2/\mathbb{Z}_2$, but non-invertible selection rules remain \cite{Kobayashi:2024yqq}.
Note that $\psi^k_{T^2/\mathbb{Z}_2}$ behave as if they have $\mathbb{Z}_M$ charges, $k$ and $-k$.
Such non-invertible selection rules are referred to as $\mathbb{Z}_2$ gauging of $\mathbb{Z}_M$ symmetry. 
They correspond to certain conjugacy classes of $D_M \simeq \mathbb{Z}_M \rtimes\mathbb{Z}_2$.\footnote{Here, we follow the group notations in Refs.~\cite{Ishimori:2010au,Kobayashi:2022moq}.}

 When we apply the above non-invertible selection rules to particle physics, we can derive novel results.
One can derive interesting textures of quarks and leptons
\cite{Kobayashi:2024cvp,Kobayashi:2025znw,Kobayashi:2025ldi,Jiang:2025psz,Qu:2026omn,Nomura:2025yoa}.
Some textures can solve the strong CP problem without introducing the axion~\cite{Liang:2025dkm,Kobayashi:2025thd,Kobayashi:2025rpx}.
The neutrino and dark matter physics were also studied in Refs.~\cite{Kobayashi:2025cwx,Nomura:2025sod,Chen:2025awz,Okada:2025kfm,Jangid:2025krp,Nomura:2025tvz,Suzuki:2025oov}.
Further applications to various directions have been studied in Refs.~\cite{Kobayashi:2025lar,Kobayashi:2025wty,Nomura:2026hcu,Nakai:2025thw}.
Thus, non-invertible selection rules have become a quite important concept in recent particle physics.

In this paper, we revisit modular flavor symmetries 
in magnetized compactifications with Scherk Schwarz (SS) phases.
We show non-invertible properties of modular symmetries.
That is, a single state is written by a linear combination of states, whose transformation behaviors under the modular symmetry are different from each other, like $\mathbb{Z}_2$ gauging of $\mathbb{Z}_M$ symmetries.
As a result, different types of modular forms appear as coupling constants.

This paper is organized as follows.
In section 2, we give a brief review on magnetized torus models and their modular symmetries.
In section 3, we study the modular symmetry of magnetized compactifications with SS phases.
Section 4 is devoted to conclusion.

\section{Magnetized torus models and modular symmetry}

In this section, we give a brief review on 
wave functions on $T^2$ with magnetic flux and 
their modular transformation properties.

\subsection{Magnetized torus models}
Let us first introduce the geometry of the two-dimensional torus. 
We use the complex coordinate $z=x+\tau y$, where $x$ and $y$ are real coordinates, while $\tau$ denotes the 
complex structure modulus.
The torus $T^2$ is defined by the identification $z \sim z + 1 \sim z+\tau$.

We consider a $U(1)$ gauge theory with the following magnetic flux background:
\begin{align}
F = \frac{\pi i M}{{\rm Im}\tau}dz \wedge d\bar z,
\end{align}
where Dirac's quantization condition requires $M$ to be an integer.
For a fermionic field $(\psi_+,\psi_-)^T$ with $U(1)$ charge $q=1$, we study the zero-modes of the Dirac equation, $i\slash{D} \psi=0$, in this magnetic flux 
background.
When $M>0$ ($M<0$), only the $\psi_+$ ($\psi_-$) modes have zero-mode solutions. 
In what follows, we restrict ourselves to the positive flux case, $M>0$.
For simplicity, we omit the subscript $+$ of $\psi_+$ and denote it just by $\psi$. 
In addition to the magnetic flux, we introduce SS phases $(\alpha_1,\alpha_2)$.
Then, the zero-mode wave functions with SS phase $(\alpha_1, \alpha_2)$ satisfy the following boundary conditions:
\begin{align}
&\psi^{j,M}_{\alpha_1,\alpha_2}(z+1,\tau) = e^{2\pi i \alpha_1}e^{\pi iM\frac{{\rm Im} z}{{\rm Im} \tau}}\psi^{j,M}_{\alpha_1,\alpha_2}(z,\tau),\\
&\psi^{j,M}_{\alpha_1,\alpha_2}(z+\tau,\tau) = e^{2\pi i \alpha_2}e^{\pi iM\frac{{\rm Im} \bar{\tau}z}{{\rm Im} \tau}}\psi^{j,M}_{\alpha_1,\alpha_2}(z,\tau).
\end{align}
Note that the phases due to the Wilson lines are equivalent to the SS phases.
(See Appendix of Ref.~\cite{Abe:2013bca}.)
We set the Wilson lines to zero.

The number of zero-modes is $M$. 
Their wave functions are denoted by $\psi^{j,M}_{\alpha_1,\alpha_2}(z,\tau)$ with $j=0,1,\cdots, M-1$, and are written as \cite{Cremades:2004wa,Abe:2013bca}
\begin{align}
    \psi^{j,M}_{\alpha_1,\alpha_2}(z,\tau) = \mathcal{N}e^{\pi iMz \frac{{\rm Im}z}{{\rm Im}\tau}}\vartheta
    \begin{bmatrix}
    \frac{j+\alpha_1}{M} \\
    -\alpha_2 \\
    \end{bmatrix}
  (Mz,M\tau),
\end{align}
where $\mathcal N$ is the normalization factor.
Here $\vartheta$ denotes the Jacobi theta function, which is defined by
\begin{align}
  \vartheta
  \begin{bmatrix}
    a \\ b \\
  \end{bmatrix}
  (\nu,\tau)
  = \sum_{\ell\in\mathbb{Z}} e^{\pi i(a+\ell)^2\tau} e^{2\pi i(a+\ell)(\nu+b)}.
\label{eq:Jacobi_theta_function}
\end{align}
The wave functions are normalized according to
\begin{align}
    \int_{T^2}dzd\bar{z}(\psi^{j,M}_{\alpha_1,\alpha_2})^{\star}\psi^{k,M}_{\beta_1,\beta_2} = (2{\rm Im}\tau)^{-1/2}\delta_{j,k}\delta_{\alpha_1,\beta_1}\delta_{\alpha_2,\beta_2}.
\end{align}

The wave functions of bosonic modes are written by the same wave functions, $\psi^{j,M}_{\alpha_1,\alpha_2}$.
These wave functions have the interesting property that their products can be expanded by other wave functions \cite{Cremades:2004wa},
\begin{align}
\label{eq:wf-expansion}
\psi^{j,M}_{\alpha_1,\alpha_2}(z,\tau)~\psi^{j',M'}_{\alpha'_1,\alpha'_2}(z,\tau)=\sum_{k,\beta_1,\beta_2} Y^{(j,M)(j'M')(k,M+M')}_{(\alpha_1,\alpha_2)(\alpha_1',\alpha'_2)(\beta_1,\beta_2)} (\tau) ~\psi^{k,M+M'}_{\beta_1,\beta_2}(z,\tau),
\end{align}
where the expansion coefficients depend on $\tau$,
\begin{align}
    Y^{(j,M)(j'M')(k,M+M')}_{(\alpha_1,\alpha_2)(\alpha_1',\alpha'_2)(\beta_1,\beta_2)} (\tau) &= \mathcal{A}^{-1/2} \left( \frac{MM'}{M+M'} \right)^{1/4} \notag \\
    & \quad \times \vartheta \begin{bmatrix} \frac{M'(j+\alpha_1) - M(j'+\alpha_1') + MM'k}{MM'(M+M')} \\ M' \alpha_2 - M \alpha_2' \end{bmatrix} \left( 0, MM'(M+M') \tau \right)  .
\end{align}
Yukawa couplings among two fermions and a boson are obtained by the overlap integral of products of wave functions,
\begin{align}
    Y^{(j,M)(j'M')(k,M+M')}_{(\alpha_1,\alpha_2)(\alpha_1',\alpha'_2)(\beta_1,\beta_2)} (\tau)=\int dz d\bar{z}~ \psi^{j,M}_{\alpha_1,\alpha_2}~\psi^{j',M'}_{\alpha'_1,\alpha'_2}~(\psi^{k,M'}_{\beta_1,\beta_2})^*,
\end{align}
up to normalization.

\subsection{Modular transformation}
The modular transformation is characterized by an element $\gamma$ acting on the modulus $\tau$ and the coordinate $z$ as
\begin{align}
\gamma : \tau \to \tau' = \frac{a \tau +b}{c\tau +d},
\end{align}
and
\begin{align}
\gamma : z \to z'=\frac{z}{c\tau +d}.
\end{align}
Here, $\gamma$ is an element of the modular group $\Gamma = SL(2,\mathbb{Z})$, which is given by a $(2 \times 2)$ matrix
\begin{align}
\gamma = 
\begin{pmatrix}
a & b\\ c & d\\
\end{pmatrix},
\end{align}
where $a,b,c,d$ are integers satisfying $ad-bc=1$.

The group $\Gamma$ is generated by two generators $S$ and $T$,
\begin{align}
S=
\begin{pmatrix}
0 & 1 \\ -1 & 0
\end{pmatrix}, 
\qquad 
T=
\begin{pmatrix}
1 & 1 \\ 0 & 1
\end{pmatrix}.
\end{align}
These generators obey the relations
\begin{align}
S^2=-1, \qquad (ST)^3=1.
\end{align}

We now discuss the modular transformation properties of the zero-mode wave functions on $T^2$.
These wave functions transform under the double covering group of $\Gamma$, denoted by $\tilde \Gamma$ 
\cite{Kikuchi:2020frp,Kikuchi:2021ogn}.
For the definition of $\tilde \Gamma$, see Ref.~\cite{Liu:2020msy}.
To represent $\tilde \Gamma$, we introduce a new element $\tilde Z$, which extends the center of $\Gamma$.
The generators $\tilde S$ and $\tilde T$ of $\tilde \Gamma$ satisfy
\begin{align}
\tilde S^2 = \tilde Z, \quad (\tilde S \tilde T)^3=\tilde Z^2, \quad \tilde Z^4=1, \quad \tilde Z \tilde T = \tilde T \tilde Z.
\end{align}

First, we show the modular transformations of wave functions with vanishing SS phases.
Under the $\tilde S$ transformation, the zero-mode wave functions are transformed according to
\begin{align}
&\psi^{j,M}_{0,0}(-z/\tau, -1/\tau) =\tilde J(\tilde S,\tau)\sum_k \rho(\tilde S)_{jk}\psi^{k,M}_{0,0}(z,\tau), \\
&\rho(\tilde S)_{jk}=e^{\pi i/4}\frac{1}{\sqrt M}e^{2\pi i \frac{jk}{M}}.
\end{align}
The factor $\tilde J(\tilde S,\tau)$ is the automorphy factor and is given by
$\tilde J(\tilde S,\tau)=(-\tau)^{1/2}$.
This implies that the zero-mode wave functions carry modular weight $1/2$. Under the $\tilde T$ transformation, the zero-modes wave functions transform as 
\begin{align}
  \psi^{j,M}(z,\tau) &\to \psi^{j,M}(z,\tau+1) \\
  &= \mathcal{N}  e^{\pi iMz\frac{{\rm Im}z}{{\rm Im}\tau}}
  \sum_{\ell\in \mathbb{Z}} e^{\pi iM\Big( l+\frac{j}{M} \Big)^2(\tau+1)}e^{2\pi iM\Big( l+\frac{j}{M} \Big)z}.
\end{align}
For even $M$, the $\tilde T$ transformation closes among $\psi^{j,M}$ with $j=0,\ldots,M-1$, and can be expressed as follows:
\begin{align}
&\psi^{j,M}_{0,0}(z, \tau + 1) =\sum_k \rho(\tilde T)_{jk}\psi^{k,M}_{0,0}(z,\tau),\\
&\rho(\tilde T)_{jk}=e^{\pi i\frac{j^2}{M}}\delta_{j,k}.
\end{align}
Note that $(\rho(\tilde{T}))^{2M}=1$, and it generates the $\mathbb{Z}_{2M}$ group. 
As a result, the modular symmetry is represented by the wave functions with vanishing SS phases, $\psi^{j,M}_{0,0}$ only if $M=$ even.
Then, the modular symmetries for models with $M=$ even were studied for vanishing SS phases.

As a result, the wave functions transform properly under the modular symmetry.
Since Yukawa couplings and the wave function satisfy the relation (\ref{eq:wf-expansion}), Yukawa couplings must also transform under the modular symmetry in the proper way consistent with Eq.~(\ref{eq:wf-expansion}).
That is, Yukawa couplings are modular forms if the modular symmetry is represented manifestly in the wave functions.
In particular, the wave functions are eigenfunctions of $T$ transformation.
Also, Yukawa couplings, $Y$ are eigenfunctions of $T$ transformations.
Usually, generic modular forms are eigenfunctions of $T$ transformations and sometimes written by $q$ expansions, where $q=e^{2\pi i \tau}$.
(See Refs.~\cite{Kobayashi:2018vbk,Feruglio:2017spp,Penedo:2018nmg,Novichkov:2018nkm,Liu:2019khw,Novichkov:2020eep,Liu:2020akv,Wang:2020lxk,Liu:2020msy} for modular form of discrete modular groups such as $S_3$, $A_4$, $S_4$, $A_5$ and their covering groups.)

Similarly, we can study the modular symmetry of wave functions including non-vanishing SS phases \cite{Kikuchi:2021ogn}.
In this case, the modular transformation properties can be written as
\begin{align}
&\psi^{j,M}_{\alpha_1,\alpha_2}(z',\tau') =\tilde J(\tilde \gamma,\tau)\sum_k\sum_{\beta_1,\beta_2} \rho(\tilde \gamma)_{(j\alpha)(k\beta)}\psi^{k,M}_{\beta_1,\beta_2}(z,\tau), \\
&\rho(\tilde S)_{(j\alpha)(k\beta)}=e^{\pi i/4}\frac{1}{\sqrt M}e^{\frac{2\pi i}{M}\{ (j+1)k+(1-\alpha_1)\beta_1 \}}\delta_{\alpha_2,\beta_1}\delta_{\beta_2,\alpha_2-\alpha_1+\frac{M}{2}},\\
&\rho(\tilde T)_{(j\alpha)(k\beta)}= e^{\frac{\pi i}{M}(j+\alpha_1)(j-\alpha_1+m)}\delta_{j,k}\delta_{\alpha_1,\beta_1}\delta_{\beta_2,\alpha_2-\alpha_1+\frac{M}{2}}.
\end{align}
Note that $m=M\pmod2,$ with $m=0,1$ and Kronecker delta associated with SS phase is understood up to integer shifts.
That implies the following transformation behaviors,
that the wave functions with SS phase $(0,0)$ transform to ones with $(0,1/2)$ under $T$ transformation when $M=$ odd, and ones with $(0,1/2)$ transform to ones with $(1/2,0)$ under $S$.
Also, the wave functions with  SS phase $(1/2,0)$ transform to ones with $(1/2,1/2)$ under $T$ transformation when $M=$ even, and ones with $(1/2,0)$ transform to ones with $(0,1/2)$.
Thus, the wave functions with the SS phases, $(0,0)$, $(0,1/2)$, $(1/2,0)$, and $(1/2,1/2)$ transform each other under the modular symmetry, in general, although the modular transformations are closed in wave functions with $(0,0)$ for $M=$ even and with $(1/2,1/2)$ for $M=$ odd.
We find that the modular symmetry is broken except in these cases.
However, we revisit this point in Section 3.

\subsection{$T^2/\mathbb{Z}_2$ orbifold models}

Next, we discuss the orbifold $T^2/\mathbb{Z}_2$, where 
the identification $z \sim -z$ is imposed.
Only the four types of SS phases are possible \cite{Abe:2013bca}, 
\begin{align}
\label{eq:SS}
    (\alpha_1,\alpha_2)=(0,0),~(0,1/2),~(1/2,0),~(1/2,1/2),
\end{align}
including the trivial one.
Then, the wave functions on $T^2/\mathbb{Z}_2$ are written by use of 
the above wave functions on $T^2$ as
\begin{align}
    \psi^{j,M}_{\pm~\alpha_1,\alpha_2}(z,\tau)={\cal N}^{(j)}\left( \psi^{j,M}_{\alpha_1,\alpha_2}(z,\tau) \pm \psi^{j,M}_{\alpha_1,\alpha_2}(-z,\tau)\right).
\end{align}
The wave functions on $T^2$ have the following property:
\begin{align}
    \psi^{j,M}_{\alpha_1,\alpha_2}(-z,\tau)=e^{-4\pi i (j+\alpha_1)\alpha_2/M }\psi^{M-j,M}_{-\alpha_1,-\alpha_2}(z,\tau).
\end{align}
Using this property, we can write \cite{Abe:2008fi,Abe:2013bca}
\begin{align}
\label{eq:wf-orbi-SS}
    \psi^{j,M}_{\pm~\alpha_1,\alpha_2}(z,\tau)={\cal N}^{(j)}\left( \psi^{j,M}_{\alpha_1,\alpha_2}(z,\tau) \pm e^{-4\pi i (j+\alpha_1)\alpha_2/M }\psi^{M-j,M}_{-\alpha_1,-\alpha_2}(z,\tau)\right).
\end{align}

For example, when $(\alpha_1,\alpha_2)=(0,0)$, the $\mathbb{Z}_2$ even wave functions on $T^2/\mathbb{Z}_2$ can be written by 
\begin{align}
    \psi^{j,M}_{+ 0,0}(z,\tau) = {\cal N}^{(j)}\left(\psi^{j,M}_{0,0}(z,\tau)+\psi^{M-j,M}_{0,0}(z,\tau)  \right).
\end{align}
The magnetized $T^2$ models have the $\mathbb{Z}_M$ symmetry, where the wave functions $\psi^{j,M}_{0,0}$ on $T^2$ transform as \cite{Abe:2009vi}
\begin{align}
    \psi^{j,M}_{0,0} \to e^{2\pi i j/M}\psi^{j,M}_{0,0}.
\end{align}
That is, $\psi^{j,M}_{0,0}$ has the $\mathbb{Z}_M$ charge $j$.
However, the wave function $\psi^{j,M}_{+ 0,0}$ on $T^2/\mathbb{Z}_2$ behaves as if it has both charges, $j$ and $M-j$.
The $\mathbb{Z}_M$ symmetry is broken, but $\mathbb{Z}_2$ gauging of $\mathbb{Z}_M$ remains as a selection rule \cite{Kobayashi:2024yqq,Kobayashi:2024cvp}.

Here, we define classes $[g^k]$ as follows,
\begin{align}
    [g^k]=\{hgh^{-1} | h=e,r   \},
\end{align}
where $g$ denotes the generator of $\mathbb{Z}_M$, i.e., $g=e^{2 \pi i/M}$, and 
\begin{align}
    eg^ke^{-1}=g^k,\qquad rg^kr^{-1}=g^{M-k}.
\end{align}
Their multiplication rules are written by 
\begin{align}
    [g^k][g^{k'}]=[g^{k+k'}]+[g^{k-k'}].
\end{align}
That is non-invertible, and corresponds to a hypergroup, in particular a fusion algebra.
The classes $[g^k]$ correspond to $\mathbb{Z}_2$ invariant conjugacy classes of $D_M \simeq \mathbb{Z}_M \rtimes \mathbb{Z}_2$.
Similarly, each string state in heterotic orbifold models corresponds not to a space group element, but to a conjugacy class of the space group \cite{Dixon:1986jc,Hamidi:1986vh,Dixon:1986qv}, which include several group elements.

Multiplication rules of (conjugacy) classes are different from those of group elements,
\begin{align}
\label{eq:xyz}
    xy=\sum_z N_{xy}^z z.
\end{align}
In group theory, 
the element appearing in the right hand side of Eq.(\ref{eq:xyz}) is unique and one can define the inverse of each element.
However, more than one elements appear in the right hand side in 
multiplication rules of conjugacy classes and more generic fusion algebras, because these classes include more than one elements.
One can not define inverses, and the algebra is non-invertible.
It is consistent to impose these multiplication rules as selection rules in field theory as imposing group-like selection rules \cite{Kaidi:2024wio}.
Indeed, field theory with non-invertible section rules lead to novel results in particle physics as mentioned in Introduction.

One of important aspects in the above discussion is as follows.
We may start with the basis of fields and string modes such that group-like symmetries are manifest.
Such symmetries are $\mathbb{Z}_M$ and $D_M$ in the above example.
We move to another basis by some reasons, e.g. geometrical requirements and boundary conditions.
The original group-like symmetries seems to be violated.
However, some "symmetries" may remain, although they may not be manifest.
These may be symmetries of subgroup,  non-invertible selection rules, or anything else.

\section{Non-invertible properties}

In this section, we revisit the modular symmetry in magnetized $T^2$ models.
We restrict ourselves to the wave functions with the SS phases $(0,0)$, $(0,1/2)$, $(1/2,0)$, $(1/2,1/2)$ as shown in Eq.~(\ref{eq:SS}), because they transform each other and these SS phases are consistent with the $\mathbb{Z}_2$ orbifolding.

\subsection{Odd flux $M$}

First, let us study the models with $M=$ odd.
Here, we write $\rho(\tilde S)$ and $\rho(\tilde T)$ in the matrix form,
\begin{equation}
\rho(\tilde T)_{(i\alpha)(j\beta)} =
\begin{pNiceArray}{cc:cc}[first-row,first-col,
                          code-for-first-col=\scriptstyle,
                          code-for-first-row=\scriptstyle,
                          columns-width = auto]
                     & j(0,0) & j(0,\tfrac12) & j(\tfrac12,0) & j(\tfrac12,\tfrac12) \\
  i(0,0)             & 0      & A_{ij}           & 0             & 0                    \\
  i(0,\frac12)       & A_{ij}    & 0             & 0             & 0                    \\ \hdashline
  i(\frac12,0)       & 0      & 0             & B_{ij}           & 0                    \\
  i(\frac12,\frac12) & 0      & 0             & 0             & B_{ij}                  \\
\end{pNiceArray},
\quad
A_{ij} \equiv \delta_{ij} e^{\frac{\pi i}{M}i(i+1)},
\quad
B_i \equiv \delta_{ij} e^{\frac{\pi i}{M} (i+\frac12)^2},
\end{equation}

\begin{equation}
\rho(\tilde S)_{(i\alpha)(j\beta)} =
\begin{pNiceArray}{cc:cc}[first-row,first-col,
                          code-for-first-col=\scriptstyle,
                          code-for-first-row=\scriptstyle,
                          columns-width = auto]
                     & j(0,0) & j(0,\tfrac12)           & j(\tfrac12,0)           & j(\tfrac12,\tfrac12)     \\
  i(0,0)             & C_{ij} & 0                       & 0                       & 0                        \\
  i(0,\frac12)       & 0      & 0                       & C_{ij} e^{\frac{\pi i}{M}} & 0                        \\ \hdashline
  i(\frac12,0)       & 0      & C_{ij} e^{\frac{\pi i}{M}} & 0                       & 0                        \\
  i(\frac12,\frac12) & 0      & 0                       & 0                       & C_{ij} e^{\frac{\pi i}{2M}} \\
\end{pNiceArray},
\quad
C_{ij} \equiv e^{i\pi/4} \frac1{\sqrt{M}} e^{\frac{2\pi i}{M} (i+1)j}.
\end{equation}
It is found that the sectors with the SS phases $(0,0)$, $(0,1/2)$, and $(1/2,0)$ transform each other.

If we neglect the phases, $\rho(\tilde T)$ and $\rho(\tilde T)$ become 
the following permutations of the sectors with the SS phases $(0,0)$, $(0,1/2)$, and $(1/2,0)$: 
\begin{align}
    \rho(t)=
    \begin{pmatrix}
        0 & 1 & 0 \\
        1 & 0 & 0 \\
        0 & 0 & 1
    \end{pmatrix}, \quad 
    \rho(s)=
    \begin{pmatrix}
        1  & 0 & 0 \\
        0  & 0 & 1 \\
        0  & 1 & 0
    \end{pmatrix}.
\end{align}
They satisfy the following algebraic relations:
\begin{align}
    (\rho (s)\rho (t))^3=(\rho(t))^2=1.
\end{align}
That is $S_3$, where 
\begin{align}
    \rho (s)\rho (t)=
\begin{pmatrix}
    0 & 1 & 0 \\
    0 & 0 & 1 \\
    1 & 0 & 0
\end{pmatrix}.
\end{align}
That implies that the modular symmetry is a permutation among the three sectors.

We focus on the $T$ transformation, because modular forms are written in the basis that $T$ is diagonal, i.e., $q=e^{2\pi i \tau}$ expansions.
The $T$ transformation is diagonalized in the following basis:
\begin{equation}\phantomsection\label{eq:basis_odd}{
\begin{cases}
  \displaystyle \Psi^{i,M}_{(0,\pm)}(z,\tau) \equiv \frac1{\sqrt{2}} \pqty{\psi^{i,M}_{(0,0)}(z,\tau) \pm \psi^{i,M}_{(0,\frac12)}(z,\tau)}, \\
  \Psi^{i,M}_{(\frac12,\alpha_2)}(z,\tau) \equiv \psi^{i,M}_{(\frac12,\alpha_2)}(z,\tau). \\
\end{cases}
}\end{equation}
Then, the $T$ transformation is represented by 
\begin{align}
    T \Psi^{i,M}_{\alpha} = e^{2\pi ih_{i\alpha}/(2M)} \Psi^{i,M}_{\alpha},
\end{align}
with $\alpha=(0,+), (0,-), (1/2,\alpha_2)$, where
\begin{align}
h_{i(0,+)} = {i(i+1)}, 
\quad
h_{i(0,-)} = {i(i+1)} +  M,
\quad
h_{i(\frac12,\alpha_2)} = {\qty(i+\frac12)^2}.
\end{align}
That is the $\mathbb{Z}_{2M}$ symmetry.

Inversely, the original wave functions with the definite SS phases are written by 
\begin{equation}
 \begin{cases}
  \displaystyle \psi^{i,M}_{(0,0)}(z,\tau) = \frac1{\sqrt{2}} \pqty{\Psi^{i,M}_{(0,+)}(z,\tau) + \Psi^{i,M}_{(0,-)}(z,\tau)}, \\
  \displaystyle \psi^{i,M}_{(0,\frac12)}(z,\tau) = \frac1{\sqrt{2}} \pqty{\Psi^{i,M}_{(0,+)}(z,\tau) - \Psi^{i,M}_{(0,-)}(z,\tau)}, \\
  \psi^{i,M}_{(\frac12,\alpha_2)}(z,\tau) = \Psi^{i,M}_{(\frac12,\alpha_2)}(z,\tau). \\
\end{cases}
\end{equation}
Note that the sectors with the SS phase $(0,0)$ and $(0,1/2)$ behave as if they have two $\mathbb{Z}_{2M}$ charges, $h_{i(0,+)}=i(i+1)$ and $h_{i(0,-)}=i(i+1)+M$.
We may understand that $\mathbb{Z}_{2M}$ is broken to $\mathbb{Z}_M$.
Indeed, $T^2$ transformation leading to $\tau \to \tau +2$ are represented in the diagonal form 
on the wave functions with the definite SS phases.
However, only the $\mathbb{Z}_M$ symmetry allows 
any liner combinations of 
$\Psi^{i,M}_{(0,+)}(z,\tau)$ and $  \Psi^{i,M}_{(0,-)}(z,\tau)$
like 
\begin{align}
 a  \Psi^{i,M}_{(0,+)}(z,\tau) + b\Psi^{i,M}_{(0,-)}(z,\tau)  .
\end{align}

When we use the basis of wave functions, $\Psi^{i,M}_\alpha$, the modular symmetry is manifest.
For example, the $S$ transformation is written by 
\begin{equation}
\rho(\tilde S)_{(i\alpha)(j\beta)} =
\left(
  \begin{array}{cc:cc}
    C_{ij}/2                           & C_{ij}/2                             &   C_{ij} e^{\frac{\pi i}{M}} / \sqrt{2} & 0                        \\
    C_{ij}/2                           & C_{ij}/2                             & - C_{ij} e^{\frac{\pi i}{M}} / \sqrt{2} & 0                        \\ \hdashline
    C_{ij} e^{\frac{\pi i}{M}} / \sqrt{2} & - C_{ij} e^{\frac{\pi i}{M}} / \sqrt{2} & 0                                    & 0                        \\
    0                                  & 0                                    & 0                                    & C_{ij} e^{\frac{\pi i}{2M}} \\
  \end{array}
\right).
\end{equation}

In this basis, we can write the expansions of wave function products by
\begin{align}
\label{eq:wf-expansion'}
\Psi^{j,M}_{\alpha}(z,\tau)~\Psi^{j',M'}_{\alpha'}(z,\tau)=\sum_{k,\beta} Y^{(j,M)(j'M')(k,M+M')}_{\alpha \alpha'\beta} (\tau) ~\Psi^{k,M+M'}_\beta(z,\tau).
\end{align}
It looks like fusion algebras, although coefficients depend on the modulus $\tau$.
In this basis, the modular symmetry is manifest and $Y^{(j,M)(j'M')(k,M+M')}_{\alpha, \alpha',\beta} (\tau) $ are modular forms.
On the other hand, we use the zero-mode wave functions with the definite boundary conditions due to the SS phases.
That is just the basis change if the wave functions with all the SS phases appear in a model.
However, in a generic model, the wave functions with all the SS phases do not appear.
Then, the full modular symmetry looks violated as group-like symmetry.
At any rate, the important point is as follows.
The modular forms  in the basis $\Psi^{j,M}_\alpha$  appear even in the basis with the definite SS phases $\psi^{j,M}_{(\alpha_1,\alpha_2)}$.
The full modular symmetry is violated as group-like symmetry, but it still controls the theory.
For example, we can write 
\begin{align}
    \psi^{j,M}_{(0,0)} \psi^{(j',M)}_{(0,0)} & =\frac12 (\Psi^{j,M}_{(0,+)}+\Psi^{j,M}_{(0,-)})
    (\Psi^{j',M}_{(0,+)}+\Psi^{j',M}_{(0,-)}) \notag \\
    &=\frac12\sum_{k,\beta}Y^{(j,M)(j',M)(k,2M)}_{(0,+)(0,+)\beta}(\tau) ~\Psi^{k,2M}_\beta(z,\tau) \notag \\
    &\quad +\frac12\sum_{l,\gamma}Y^{(j,M)(j',M)(l,2M)}_{(0,+)(0,-)\gamma}(\tau)~\Psi^{l,2M}_{\gamma}(z,\tau) \notag \\
    &\quad +\frac12\sum_{m,\delta}Y^{(j,M)(j',M)(m,2M)}_{(0,-)(0,+)\delta}(\tau)~\Psi^{m,2M}_{\delta}(z,\tau) \notag \\
    &\quad +\frac12\sum_{n,\epsilon}Y^{(j,M)(j',M)(n,2M)}_{(0,-)(0,-)\epsilon}(\tau)~\Psi^{n,2M}_{\epsilon}(z,\tau)    .
    \end{align}
It is remarkable that the modular forms for the full modular symmetry appear in the right hand side.
The $T$ transformation is represented on $\Psi^{j,M}$ as $\mathbb{Z}_{2M}$.
The wave function $\psi^{j,M}_{(0,0)}$ behaves as if it has both $\mathbb{Z}_{2M}$ charges, $h_{j(0,+)}$ and  $h_{j(0,-)}$.
Thus, the right hand side includes the terms with $\mathbb{Z}_{2M}$ charges, 
$h_{j(0,\pm)}+h_{j'(0,\pm)}$.
That is non-invertible in a sense.
One may think that the $\mathbb{Z}_{2M}$ symmetry is violated to $\mathbb{Z}_M$, but coefficients $Y$ are modular forms of the full symmetry.
Hence, the full modular symmetry controls the theory.
As mentioned above, note that the wave functions with all the SS phases not appear in a model, but it depends on a model.
Incomplete multiplet representations appear.

Since the above discussions are consistent with $\mathbb{Z}_2$ orbifolding, we can study the magnetized $T^2/\mathbb{Z}_2$ orbifold models with SS phases in a similar way.

\subsection{Even flux $M$}

Similarly, we can study the models with $M=$ even.
Here, we write $\rho(\tilde S)$ and $\rho \tilde (T)$ in the matrix form,
\begin{equation}
\rho(\tilde T)_{(i\alpha)(j\beta)} =
\begin{pNiceArray}{cc:cc}[first-row,first-col,
                          code-for-first-col=\scriptstyle,
                          code-for-first-row=\scriptstyle,
                          columns-width = auto]
                     & j(0,0) & j(0,\tfrac12) & j(\tfrac12,0) & j(\tfrac12,\tfrac12) \\
  i(0,0)             & A'_{ij}        & 0                 & 0             & 0                    \\
  i(0,\frac12)       & 0             & A'_{ij}            & 0             & 0                    \\ \hdashline
  i(\frac12,0)       & 0         & 0             & 0         & B'_{ij}           \\
  i(\frac12,\frac12) & 0         & 0             & B'_{ij}    & 0                \\
\end{pNiceArray},
\quad
A'_{ij} \equiv \delta_{ij} e^{\frac{\pi i}{M}i^2},
\quad
B'_i \equiv \delta_{ij} e^{\frac{\pi i}{M} (i+\frac12)(i-\frac12)},
\end{equation}
with $\rho(\tilde S)$ unchanged from the odd case.
It is found that the sectors with the SS phases $(0,1/2)$, $(1/2,0)$, and $(1/2,1/2)$ transform each other.
If we neglect the phases, 
the permutation matrices among the SS phases $(0,1/2)$, $(1/2,0)$, and $(1/2,1/2)$ are
\begin{align}
    \rho(t)=
    \begin{pmatrix}
        1  & 0 & 0 \\
        0  & 0 & 1 \\
        0  & 1 & 0
    \end{pmatrix}. \quad 
    \rho(s)=
    \begin{pmatrix}
        0 & 1 & 0 \\
        1 & 0 & 0 \\
        0 & 0 & 1
    \end{pmatrix},
\end{align}
satisfying the same $S_3$ relations $(\rho (s)\rho (t))^3=(\rho(t))^2=1$ where 
\begin{align}
    \rho (s)\rho (t)=
    \begin{pmatrix}
        0 & 0 & 1 \\
        1 & 0 & 0 \\
        0 & 1 & 0
    \end{pmatrix}.
\end{align}

The $T$-diagonalized basis is now
\begin{equation}\phantomsection\label{eq:basis_even}{
\begin{cases}
  \displaystyle \Psi^{i,M}_{(0,\pm)}(z,\tau) \equiv \psi^{i,M}_{(0,\alpha_2)}(z,\tau), \\
  \Psi^{i,M}_{(\frac12,\alpha_2)}(z,\tau) \equiv \frac1{\sqrt{2}} \pqty{\psi^{i,M}_{(\frac12,0)}(z,\tau) \pm \psi^{i,M}_{(\frac12,\frac12)}(z,\tau)}. \\
\end{cases}
}\end{equation}
which is the counterpart of \eqref{eq:basis_odd} with the roles of $(0,\pm)$ and $(\frac12,\pm)$ exchanged.
The $T$ eigenvalues are
\begin{align}
h_{i(0,\alpha_2)} = {i^2}, 
\quad
h_{i(\frac12,+)} = {\qty(i+\frac12)\qty(i-\frac12)},
\quad
h_{i(\frac12,-)} = {\qty(i+\frac12)\qty(i-\frac12) + M},
\end{align}
with $T\Psi^{i,M}_\alpha = e^{2\pi i h_{i\alpha}/(2M)}\Psi^{i,M}_\alpha$ as before.
The $(\frac12,0)$ and $(\frac12,\frac12)$ sectors behave as if they carry two $\mathbb{Z}_{2M}$ charges, and the $\mathbb{Z}_{2M}$ symmetry seems to be broken to $\mathbb{Z}_M$ on the definite SS phase basis
\begin{equation}
 \begin{cases}
  \psi^{i,M}_{(0,\alpha_2)}(z,\tau) = \Psi^{i,M}_{(0,\alpha_2)}(z,\tau), \\
  \displaystyle \psi^{i,M}_{(\frac12,0)}(z,\tau) = \frac1{\sqrt{2}} \pqty{\Psi^{i,M}_{(\frac12,+)}(z,\tau) + \Psi^{i,M}_{(\frac12,-)}(z,\tau)}, \\
  \displaystyle \psi^{i,M}_{(\frac12,\frac12)}(z,\tau) = \frac1{\sqrt{2}} \pqty{\Psi^{i,M}_{(\frac12,+)}(z,\tau) - \Psi^{i,M}_{(\frac12,-)}(z,\tau)}. \\
\end{cases}
\end{equation}

The $S$ transformation in this basis is
\begin{equation}
\rho(\tilde S)_{(i\alpha)(j\beta)} =
\left(
  \begin{array}{cc:cc}
    C_{ij} & 0                                       & 0                                       & 0                        \\
    0      & 0                                       & - C_{ij} e^{\frac{\pi i}{M}} / \sqrt{2} & - C_{ij} e^{\frac{\pi i}{M}} / \sqrt{2}                        \\ \hdashline
    0      & - C_{ij} e^{\frac{\pi i}{M}} / \sqrt{2} & C_{ij} e^{\frac{\pi i}{2M}} / 2 & C_{ij} e^{\frac{\pi i}{2M}} / 2 \\
    0      & C_{ij} e^{\frac{\pi i}{M}} / \sqrt{2}   & C_{ij} e^{\frac{\pi i}{2M}} / 2 & C_{ij} e^{\frac{\pi i}{2M}} / 2 \\
  \end{array}
\right).
\end{equation}
The wave function product expansion and the role of modular forms are entirely analogous to the odd-$M$ case \eqref{eq:wf-expansion'}, with the $(\frac12,0)$-sector product providing a representative example:
\begin{align}
    \psi^{j,M}_{(\frac12,0)} \psi^{j',M'}_{(0,0)} 
    &=\frac{1}{2}\sum_{k,\beta}Y^{(j,M)(j',M')(k,M+M')}_{(\frac12,+)(\frac12,+)\beta}(\tau)\,\Psi^{k,M+M'}_\beta \notag \\
    &\quad+\frac{1}{2}\sum_{l,\gamma}Y^{(j,M)(j',M')(l,M+M')}_{(\frac12,+)(\frac12,-)\gamma}(\tau)\,\Psi^{l,M+M'}_{\gamma} \notag \\
    &\quad+\frac{1}{2}\sum_{m,\delta}Y^{(j,M)(j',M')(m,M+M')}_{(\frac12,-)(\frac12,+)\delta}(\tau)\,\Psi^{m,M+M'}_{\delta} \notag \\
    &\quad+\frac{1}{2}\sum_{n,\epsilon}Y^{(j,M)(j',M')(n,M+M')}_{(\frac12,-)(\frac12,-)\epsilon}(\tau)\,\Psi^{n,M+M'}_{\epsilon}.
\end{align}
Note that which SS-phase sectors actually appear depends on the specific model.
Incomplete multiplet representations appear.

Since the above discussions are consistent with $\mathbb{Z}_2$ orbifolding, we can study the magnetized $T^2/\mathbb{Z}_2$ orbifold models with SS phases in a similar way.

\subsection{Examples}

Here, we give examples for smaller $M$.
First, 
for $M=1$, $\rho(\tilde S)$ and $\rho(\tilde T)$ (where $\omega \equiv e^{\pi i/4}$) are represented by 
\begin{equation}
\rho(\tilde S)_{(i\alpha)(j\beta)} =
\left(
  \begin{array}{c:c:c:c}
    \omega &    &    &    \\ \hdashline
      &    & -\omega &    \\ \hdashline
      & -\omega &    &    \\ \hdashline
      &    &    & i\omega \\
  \end{array}
\right),
\end{equation}
\begin{equation}
\rho(\tilde T)_{(i\alpha)(j\beta)} =
\left(
  \begin{array}{c:c:c:c}
      & 1 &   &   \\ \hdashline
    1 &   &   &   \\ \hdashline
      &   & \omega &   \\ \hdashline
      &   &   & \omega \\
  \end{array}
\right),
\end{equation}
in the wave function basis with definite SS phases, $\psi^{j,M}_{(\alpha_1,\alpha_2)}$.
Thus, the wave functions with the SS phases $(0,0)$, $(0,1/2)$, and $(1/2,0)$ transform each other, while the wave function with the SS phase $(1/2,1/2)$ is a singlet.

Here, we focus on the representations only on the wave functions with the SS phases $(0,0)$, $(0,1/2)$, and $(1/2,0)$, i.e. the upper left $(3 \times 3)$ submatrices of $\rho(\tilde S)$ and $\rho (\tilde T)$.
Their closed algebra is isomorphic to $\Delta(48) \rtimes \mathbb{Z}_8$, whose order is 384.
The wave functions with the SS phases $(0,0)$, $(0,1/2)$, and $(1/2,0)$ correspond to its triplet.

For $M=2$
$\rho(\tilde S)$ and $\rho(\tilde T)$ are represented by 
\begin{equation}
\scriptsize
\rho(\tilde S)_{(i\alpha)(j\beta)} =
\left(
  \begin{array}{cc:cc:cc:cc}
    \omega/\sqrt{2} & - \omega/\sqrt2 &           &             &           &             &          &            \\
    \omega/\sqrt{2} &   \omega/\sqrt2 &           &             &           &             &          &            \\ \hdashline
               &            &           &             & i\omega/\sqrt2 & - i\omega/\sqrt2 &          &            \\
               &            &           &             & i\omega/\sqrt2 &   i\omega/\sqrt2 &          &            \\ \hdashline
               &            & i\omega/\sqrt2 & - i\omega/\sqrt2 &           &             &          &            \\
               &            & i\omega/\sqrt2 &   i\omega/\sqrt2 &           &             &          &            \\ \hdashline
               &            &           &             &           &             & i/\sqrt2 & - i/\sqrt2 \\
               &            &           &             &           &             & i/\sqrt2 &   i/\sqrt2 \\
  \end{array}
\right),
\end{equation}
\begin{equation}
\rho(\tilde T)_{(i\alpha)(j\beta)} =
\left(
  \begin{array}{cc:cc:cc:cc}
    1 &   &   &   &          &         &          &           \\
      & i &   &   &          &         &          &           \\ \hdashline
      &   & 1 &   &          &         &          &           \\
      &   &   & i &          &         &          &           \\ \hdashline
      &   &   &   &          &         & \omega^{-1/2} &           \\
      &   &   &   &          &         &          & i\omega^{-1/2} \\ \hdashline
      &   &   &   & \omega^{-1/2} &         &          &           \\
      &   &   &   &          & \omega^{3/2} &          &           \\
  \end{array}
\right),
\end{equation}
in the wave function basis with definite SS phases, $\psi^{j,M}_{(\alpha_1,\alpha_2)}$.
Thus, the wave functions with the SS phases $(0,1/2)$, $(1/2,0)$, and $(1/2,1/2)$ transform each other, while the wave functions with the SS phase $(0,0)$ transform inside themselves.

Here, we focus on the representations only on the wave functions with the SS phases $(0,1/2)$, $(1/2,0)$, and $(1/2,1/2)$ i.e. the lower right $(3 \times 3)$ submatrices of $\rho(\tilde S)$ and $\rho (\tilde T)$.
Their closed algebra is isomorphic to $(\mathbb{Z}_8 \times \mathbb{Z}_8 \times ((\mathbb{Z}_4 \times \mathbb{Z}_2) \rtimes \mathbb{Z}_2)) \rtimes S_3$, whose order is 6144.
The wave functions with the SS phases $(0,1/2)$, $(1/2,0)$, and $(1/2,1/2)$ correspond to its triplets.

We can study the model with $M=3$.
Similar to the model with $M=1$, the wave functions with the SS phases $(0,0)$, $(0,1/2)$, and $(1/2,0)$ transform each other. 
Their group is very large, and its order is larger than 10,000.
The larger fluxes $M$ lead to much larger groups.

\subsection{Comments on localized modes}

Similarly, we can study the modular symmetry in magnetized $T^2/\mathbb{Z}_2$ orbifold models by use of wave functions (\ref{eq:wf-orbi-SS}).
So far, we have studied bulk modes.
Here, we comment on localized modes.
For example, there are four fixed points, $z=(0,0)$, $(0,\tau/2)$, $(1/2,0)$, and $(1/2,\tau/2)$ on the $T^2/\mathbb{Z}_2$ orbifold.
Localized magnetic fluxes at these fixed points induce localize modes \cite{Lee:2003mc,Buchmuller:2015eya,Buchmuller:2018lkz,Abe:2020vmv,Kobayashi:2022tti}.
In Ref.~\cite{Kobayashi:2024hkk}, it was shown that when localized fluxes are degenerate at all the fixed points, localized modes transform each other under the modular symmetry, and the $S$ and $T$ transformations are represented by
\begin{align}
    \rho(S)=
    \begin{pmatrix}
        e^{3\pi i \ell/2} & 0 & 0 & 0  \\
        0 & 0 & e^{\pi i \ell/2} & 0 \\
        0 & e^{\pi i \ell/2} & 0 & 0 \\
        0 & 0& 0 & e^{\pi i \ell/2}
    \end{pmatrix}, \qquad 
    \rho(T)=
    \begin{pmatrix}
        e^{\pi i \ell/2} & 0 & 0 & 0 \\
        0 &  e^{\pi i \ell/2} & 0 & 0 \\
        0 & 0 & 0 & 1 \\
        0 & 0 & 1 & 0
    \end{pmatrix},
\end{align}
where $\ell $ is an integer 
corresponding to the localized flux.
If we can resolve their degeneracy, e.g. by projecting out some modes or make some modes massive, the multiplet becomes incomplete and the modular symmetry looks violated.
However, the full modular symmetry would control the coupling terms of remaining modes.
We would study elsewhere how to realize such a situation.

Similarly, twisted string sectors at  orbifold fixed points transform in heterotic orbifold models and they are multiplets of the modular flavor symmetries \cite{Ferrara:1989qb,Lerche:1989cs,Lauer:1989ax,Lauer:1990tm}.
Their degeneracy is resolved by introducing discrete Wilson lines \cite{Ibanez:1986tp}.
It would be interesting to study the modular symmetry in such heterotic orbifold models, although the $T$ transformation is represented by a diagonal matrix \cite{Ferrara:1989qb,Lerche:1989cs,Lauer:1989ax,Lauer:1990tm}.

\subsection{Toward phenomenological application}

The previous discussions suggest a new approach to modular flavor models.
The groups shown in the previous subsections are very large, and it is complicated to show directly their applications.
For illustration, we use  models with smaller groups as toy models.

First, we start with $S_3$.
Its $\rho(S)$ and $\rho(T)$ can be represented by 
\begin{align}
\label{eq:S3}
    \rho(S)=\frac12
    \begin{pmatrix}
        -1 & -\sqrt{3} \\
        -\sqrt{3} & 1
    \end{pmatrix}, \qquad
    \rho(T) = 
    \begin{pmatrix}
        1 & 0 \\
        0 & -1
    \end{pmatrix},
\end{align}
on a doublet, and they satisfy $\rho(S)^2=(\rho(S) \rho(T))^3=\rho(T)^2=\mathbb{I}$.
The doublet corresponds to the $\Psi_\alpha$ with $\alpha = 0,1$.
However, the physical basis may correspond to 
\begin{align}
    \psi_\pm = \Psi_0 \pm \Psi_1,
\end{align}
up to normalization.
That is just the basis change in $S_3$.
However, both $\psi_+$ and $\psi_-$ do not always appear in a generic model.
In this sense, $S_3$ is violated, but 
couplings in models are controlled by $S_3$.
For example, the coupling term among $\psi_+$ and two $S_3$ trivial singlets $\psi_{\mathbb{I}}$ can be written by
\begin{align}
    (Y_1(\tau) + Y_2(\tau)) \psi_+  \psi_{\mathbb{I}} \psi_{\mathbb{I}},
\end{align}
where $(Y_1(\tau),Y_2(\tau))^T$ is a $S_3$ doublet modular form in the basis (\ref{eq:S3}).
Note that more than one modular forms appear in the coupling.

Similarly, we can  study $S_4$ models.
Its $\rho(S)$ and $\rho(T)$ can be represented by 
\begin{align}
    \rho(S)=\frac12
    \begin{pmatrix}
        0 & -\sqrt{2} & -\sqrt{2} \\
        -\sqrt{2} & -1 & 1 \\
        -\sqrt{2} & 1 & -1
    \end{pmatrix}, \qquad
    \rho(T) = 
    \begin{pmatrix}
        1 & 0 & 0 \\
        0 & i & 0 \\
        0 & 0 & -i
    \end{pmatrix},
\end{align}
on a triplet, and the satisfy $\rho(S)^2=(\rho(S) \rho(T))^3=\rho(T)^4=\mathbb{I}$.
The triplet corresponds to the $\Psi^{j,M}_\alpha$ with $\alpha=1,2,3$.
The physical basis may correspond to 
\begin{align}
    \psi_1=\Psi_1, \qquad \psi_\pm = \Psi_2 \pm \Psi_3,
\end{align}
up to normalization.
This is just the basis change in $S_4$. 
However, all of $\psi_1$ and $\psi_\pm$ do not appear in a generic model.
In this sense, $S_4$ is violated, but couplings in models are controlled by $S_4$.
For example, the coupling term among $\psi_+$ and two $S_4$ trivial singlets $\psi_{\mathbb{I}}$ can be written by 
\begin{align}
    (Y_{\bar 2}(\tau )  + Y_{\bar 3}(\tau))\psi_+\psi_{\mathbb{I}} \psi_{\mathbb{I}},
\end{align}
where $(Y_{\bar 1}(\tau),Y_{\bar 2}(\tau),Y_{\bar 3}(\tau))^T$ denote modular forms corresponding to  a $S_4$ conjugate triplet.
Note that more than one modular forms appear in the coupling.

We can discuss other modular flavor models similarly.
Our approach to modular flavor models makes model building more rich.
Such studies would be interesting, but that is beyond our scope.
We leave it as a future work.

\section{Conclusions}

We have studied the modular symmetry in magnetized compactifications with SS phases.
In general, the modular symmetry transforms the modes with different SS phases each other.
However, the modes with all the SS phases do not appear in a generic model.
Incomplete multiplet representations appear.
In this sense, the full modular symmetry is violated as a group-like symmetry.
However, the couplings are controlled by the full modular symmetry, and modular forms of the full symmetry appear as coupling constants.
That is non-trivial aspects.
When we apply our results to modular flavor models, we may construct novel models.
We leave it as a future work.

Non-invertible selection rules due to hypergroups and fusion algebras are violated by loop effects \cite{Heckman:2024obe,Kaidi:2024wio,Funakoshi:2024uvy,Dong:2026crl}.
It is important to study loop effects on non-invertible properties shown in this paper.
We leave it as a further work.

\acknowledgments

This work was supported by JSPS KAKENHI Grant Numbers JP23K03375 (T.K.), and JST SPRING, Grant Number JPMJSP2119 (R.N. and H.U.).

\bibliography{references}{}
\bibliographystyle{JHEP} 
\end{document}